\documentclass[aps,twocolumn,amsmath,amssymb,showpacs,prl]{revtex4}
\usepackage{epsf}
\usepackage{graphicx}

\newcommand{\etal}{{\it et al.}}

\begin{document}

\title{The change of Fermi surface topology in Bi$_2$Sr$_2$CaCu$_2$O$_{8+\delta}$ with doping}
\author{
        A. Kaminski,$^{1}$
       S. Rosenkranz,$^{2}$
        H. M. Fretwell,$^{1}$
        M. R. Norman,$^{2}$
        M. Randeria,$^{3}$
        J. C. Campuzano,$^{2,4}$
        J-M. Park,$^1$
        Z. Z. Li$^{5}$ and
        H. Raffy$^{5}$
      }
\affiliation{
         (1) Ames Laboratory and Department of Physics and Astronomy, Iowa State University,
             Ames, IA 50011\\
        (2) Materials Science Division, Argonne National Laboratory,
            Argonne, IL 60439 \\
         (3) Department of Physics, Ohio State University, Columbus, OH 43210\\
         (4) Department of Physics, University of Illinois at Chicago,
             Chicago, IL 60607\\
          (5) Laboratorie de Physique des Solides,
                  Universite Paris-Sud, 91405 Orsay Cedex, France\\
          }
\date{\today}
\begin{abstract}
We report the observation of a change in Fermi surface topology of Bi$_2$Sr$_2$CaCu$_2$O$_{8+\delta}$ with doping. By collecting high statistics ARPES data from moderately and highly overdoped samples and dividing the data by the Fermi function, we answer a long standing question about the Fermi surface shape of Bi$_2$Sr$_2$CaCu$_2$O$_{8+\delta}$ close to the ($\pi$,0) point. For moderately overdoped samples (T$_c$=80K) we find that both the bonding and antibonding sheets of the Fermi surface are hole-like. However for a doping level corresponding to T$_c$=55K we find that the antibonding sheet becomes electron-like. This change does not directly affect the critical temperature and therefore the superconductivity. However, since similar observations of the change of the topology of the Fermi surface were observed in LSCO \cite{FUJIMORI,INO} and Bi$_2$Sr$_2$Cu$_2$O$_{6+\delta}$ \cite{TSUNAHIRO}, it appears to be a generic feature of hole-doped superconductors. Because of bilayer splitting, though, this doping value is considerably lower than that for the single layer materials, which again argues that it is unrelated to T$_c$.
\end{abstract}

\maketitle
The Fermi surface is a fundamental property in condensed matter physics. In recent years there has been a lively debate about the shape of the Fermi surface of one of the most frequently studied high temperature superconductors - Bi$_2$Sr$_2$CaCu$_2$O$_{8+\delta}$ \cite{SHENREVIEW,JCREVIEW}. Initially, the Fermi surface of this compound was determined by angle resolved photoemission spectroscopy (ARPES) to be hole-like, and consistent with Luttinger's theorem \cite{HONGFS}. Later there were reports of an electron-like Fermi surface when measurements were performed at higher photon energies \cite{FENGFS,CHUANG}. High momentum resolution studies utilizing new generation electron analyzers re-established the hole-like shape of the Fermi surface for optimal doping \cite{HELENFS,FINKFS}. The observation of bilayer splitting in the overdoped regime \cite{FENGBILAYER} provided further evidence that at least the bonding sheet of the Fermi surface is hole-like for a wide range of dopings. However, this study did not address the question of the shape of the antibonding sheet of the Fermi surface. For a long time, the answer to this question remained elusive, as the antibonding band near ($\pi$,0) is located very close to the chemical potential. It is important to determine the exact shape of the whole Fermi surface, as it potentially affects transport and collective properties. Knowledge of the band dispersion in this region of the Brillouin zone is also of great importance for theoretical calculations. In a broader scope, it was previously shown \cite{FUJIMORI,TSUNAHIRO} that a change in the topology of the Fermi surface occurs in overdoped LSCO and heavily overdoped single layer Bi$_2$Sr$_2$CuO$_{6+\delta}$, therefore this phenomenon may be a generic property of hole-doped cuprates. Here, we present high resolution ARPES data measured to a high degree of statistical accuracy, in order to determine the evolution of the Fermi surface with doping. 
We find that for moderately doped samples, both sheets of the Fermi surface are hole-like, while at higher doping levels the antibonding band becomes electron-like. The observed change of the topology of the Fermi surface of Bi$_2$Sr$_2$CaCu$_2$O$_{8+\delta}$ occurs roughly for a similar range of doping as one reported for LSCO. 
This change is not accompanied by any abrupt change in the critical temperature at this particular value of doping. Our results resemble the effect of strain on the shape of the Fermi surface in thin films of LSCO \cite{ONELLION,BOZOVIC}. One can therefore speculate that the effects of the strain on the properties of the cuprates are very similar in nature to changes of carrier concentration.

The samples were mounted with $\Gamma-M$ parallel to the photon polarization and cleaved in situ at pressures
of less than 2$\cdot$10$^{-11}$ Torr. Measurements were carried out at the
Synchrotron Radiation Center in Madison, Wisconsin, on the U1 undulator
beamline supplying $10^{12}$ photons/sec. A Scienta SES 50 electron analyzer with an energy resolution of 20 meV and momentum resolution of 0.01 $\AA^{-1}$ for a photon energy of 22 eV. The value of the chemical potential was determined by measuring the Fermi edge of a gold film evaporated on to silicon in electrical and thermal contact with the sample. This value was stable to within 0.5 meV as verified by periodic measurements of the gold reference throughout the experiment.

\begin{figure}
\includegraphics[width=3.5in]{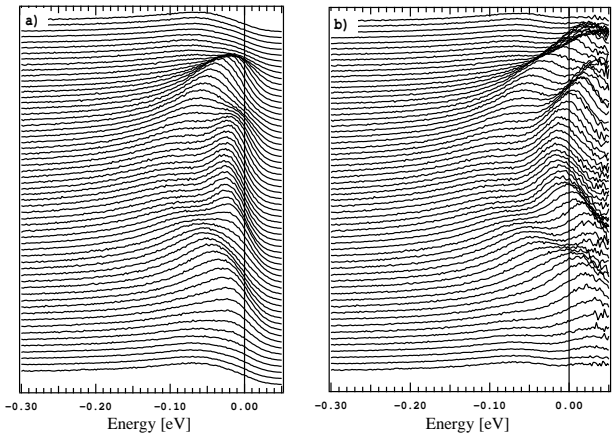}
\caption{Illustration of the division by the Fermi function method at T=100K for an overdoped sample (T$_c$=55K) along the cut (0.8$\pi,-0.3\pi$) - (0.8$\pi,0.3\pi$). a) raw ARPES data. b) data from a) divided by the resolution broadened Fermi function.}
\label{fig1}
\end{figure}

One of the limitations of the ARPES technique is the fact that it only measures the occupied part of the spectrum. The Fermi cut-off makes it difficult to directly analyze the data close to the chemical potential. To combat these limitations, we have exploited a method of gaining information about the states close to and above the chemical potential, which relies on dividing the ARPES data by the resolution broadened Fermi function \cite{SATO01,ADAMCROSS}. This function is easily obtained by fitting the gold reference spectra with a Fermi function. We renormalize the fitted Fermi function, so that its maximum amplitude is equal to 1 and then divide by it the ARPES spectra for the sample. The result is equivalent to the spectral function A(k,$\omega$) modulo matrix elements (which are almost constant within a small range of momenta). The effectiveness of this method depends on two conditions: the sample temperature and the statistics of the data. Working at high temperature increases the spectral weight above the chemical potential, and thus allows us to extract the data there with less noise. For low temperatures, data of high statistics are required, as the noise at energies above the chemical potential is rapidly amplified since the data are divided by the small values of the Fermi function. Given the high statistics of our data, we can reliably extract the spectral function for samples at T=100K up to energies of $\sim$ 60 meV above the chemical potential.  Stability of the chemical potential is essential for successful application of this method. This was monitored frequently during the experiment and the typical variation of its value was very small, of order 0.5 meV.
Another experimental consideration is the contribution to the measured electron intensity from the second order light, which is always present in synchrotron beamlines. In the case of our beamline this contribution is very small (~2\%). We determine this contribution by looking at the average intensity far above the chemical potential and then subtracting it from the data before performing the division operation.

We illustrate the division by the Fermi function in Fig. 1. In panel (a) the raw ARPES data (measured in the form of energy distribution curves or EDCs) along the ($\pi$,-$\pi $)-($\pi$,$\pi$) direction are shown, and in panel (b) the result of the division of the Fermi function is plotted. In the raw data there are two clearly visible features: (i) dispersing peaks corresponding to the bonding band and (ii) ``nondispersive" peaks near the chemical potential corresponding to the antibonding band. As stated above, it is difficult to determine the band position and dispersion of the antibonding band from such data. In contrast, it is quite easy to do so when the data are divided by the Fermi function, as shown in panel (b). Here the dispersion of both the bonding and antibonding bands are clearly visible at, and even above, the chemical potential. Such an approach allows us to determine the precise location of the bottom of both bands even in close proximity to the chemical potential.

\begin{figure}
\includegraphics[width=3.5in]{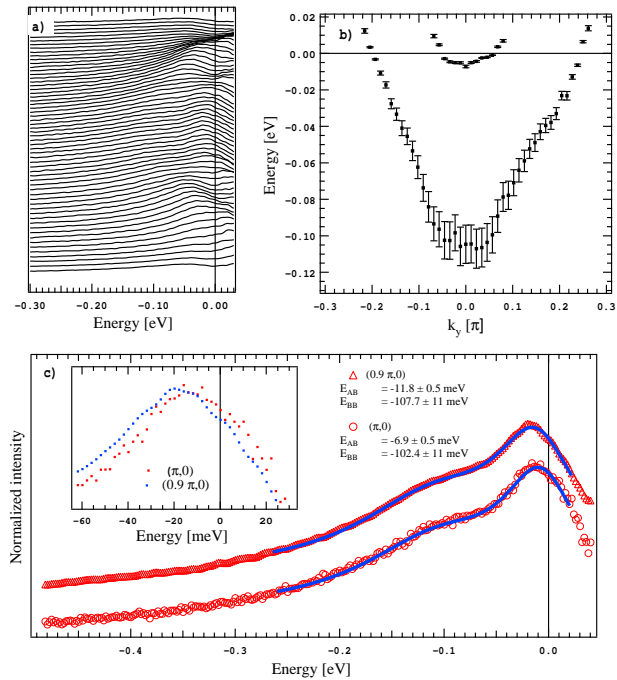}
\caption{Data along the ($\pi$,-$\pi$) to ($\pi$,$\pi$) direction for a moderately overdoped sample of T$_c$=80K at T=100K. a) EDC data divided by the Fermi function. b) dispersion  extracted from a two Lorentzian fit to the EDCs. c) EDC at ($\pi$,0) and (0.9 $\pi$,0) divided by the Fermi function (red circles and triangles) and a two Lorentzian fit (blue lines).
E$_{AB}$ and E$_{BB}$ are the positions of the antibonding and bonding peaks.
}
\label{fig2}
\end{figure}

First we examine the case of a moderately overdoped sample (T$_c$=80K). In Fig. 2 we plot the ARPES data divided by the Fermi function along the ($\pi$,-$\pi$) - ($\pi$,$\pi$) direction. The first panel shows the divided EDC curves in the proximity of the ($\pi$,0) point, while in panel (b), we show the peak dispersions for both bonding and antibonding bands extracted from the EDC data by fitting them with a sum of two Lorentzians. 
The deviations from the parabolic dispersion originate from the fact that the two Lorentzian fitting procedure becomes unstable once the antibonding peak is far above the chemical potential. This does not affect our conclusions, since we are interested in the energy position of the bottom of the bands, where this fitting procedure works very well. We note that the bottoms of both bands along this direction lie below the chemical potential, which means that for this doping value both sheets of the Fermi surface remain hole-like. To quantify this we plot in panel (c) the EDCs  at the ($\pi$,0) and (0.9$\pi$,0) points and fit them using a combination of two Lorentzian functions, corresponding to the bonding and antibonding peaks. The  ($\pi$,0) point  is at -7 meV and -102 meV for the antibonding and bonding bands respectively.

\begin{figure}
\includegraphics[width=3.5in]{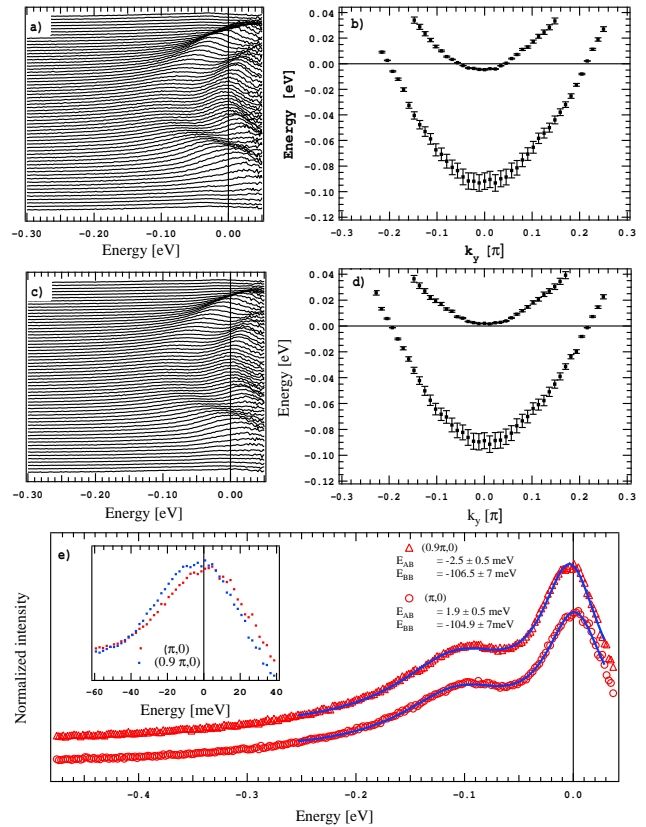}
\caption{Data along the ($\pi$,-$\pi$) to ($\pi$,$\pi$) direction for a heavily overdoped sample of T$_c$=55K at T=100K. a) EDC data divided by the Fermi function  along the (0.9$\pi$,-0.3$\pi$) - (0.9$\pi$,0.3$\pi$) cut. b) dispersion  extracted from a two Lorenzian fit to the EDC data shown in a). c) EDC data divided by the Fermi function for ($\pi$,-0.3$\pi$) - ($\pi$,0.3$\pi$) cut. d) dispersion  extracted from a two Lorentzian fit to the EDC data shown in c).  e) EDCs at ($\pi$,0) and  (0.9$\pi$,0) divided by the Fermi function (red circles) and a two Lorentzian fit (blue lines).
}
\label{fig3}
\end{figure}

Next, we turn to the most heavily overdoped sample (T$_c$=55K) that we could obtain. The data for this sample are shown in Fig.~3. In panels (a) and (b) we show the EDCs  and extracted peak dispersion along  the (0.9$\pi$,-0.3$\pi$) - (0.9$\pi$,0.3$\pi$) cut. As in the previous case, these data show that the bottoms of both bands lie below the chemical potential. However the data obtained along the Brillouin zone boundary (i.e. along ($\pi$,-0.3$\pi$) - ($\pi$,0.3$\pi$) cut), shown in panels (c) and (d), is very different. Here, the bottom of the antibonding band is located slightly above the chemical potential. To check the exact location of the antibonding band we plot the EDC curves at  ($\pi$,0) and (0.9$\pi$,0) in panel (e). Each EDC is fit with the sum of two Lorentzian functions to precisely determine the energy locations of the peaks. While the antibonding peak at (0.9$\pi$,0) is located at 2.5 meV below the chemical potential (consistent with the conclusions from data in panels (a) and (b)), the same peak at ($\pi$,0) lies at 2 meV above the chemical potential. This is clear evidence that the antibonding Fermi surface closes before reaching the zone boundary, and therefore becomes electron-like. The Fermi crossing along (0,0) - ($\pi$,0) occurs close to (0.95$\pi$,0) based on linear interpolation.

The extraction of the Fermi surface contours from the data using the standard method of plotting the intensity within a small energy range about the chemical potential over the whole Brillouin zone is quite difficult in this case. The main reasons for this are the presence of a background \cite{ADAMBACK} and the relatively wide EDC peaks in the normal state combined with the close proximity to the chemical potential of the antibonding band at ($\pi$,0). To better illustrate the topologies, in Fig.~4 we plot Fermi surfaces resulting from tight binding fits to the band dispersion extracted from the data. Panel (a) shows the Fermi surface contour for the moderately overdoped sample of T$_c$=80K. For this and lower doping levels, both sheets of the Fermi surface are hole-like, as shown by previous studies. In the case of the heavily overdoped sample (T$_c$=55K), the bonding sheet of the Fermi surface remains hole-like, but the antibonding Fermi surface becomes electron like and closes before it reaches the edge of the Brillouin zone - as illustrated in panel (b). 
\begin{figure}
\includegraphics[width=3.5in]{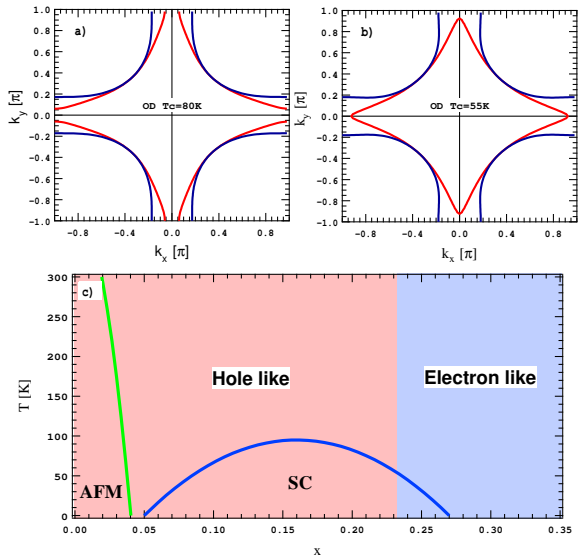}
\caption{Fermi surface resulting from a tight binding fit to data through the Brillouin zone. a) for moderately overdoped samples with T$_c$=80K. b) for heavily overdoped samples with T$_c$=55K. c) phase diagram with marked topology of the antibonding band.}
\label{fig4}
\end{figure}

The finding that the Fermi surface topology changes below the level of doping corresponding to T$_c$=55K has several important consequences. As the change of the topology is not associated with a rapid change of T$_c$, our results are not in support of those models of superconductivity which depend on the presence of a van Hove singularity. The relation of our finding to the question of quantum critical points in the phase diagram is a more subtle one.  Anomalies in specific heat data, that have been interpreted as due to the pseudogap closing, occur at a doping value of 0.19 in this material \cite{LORAM}, considerably lower than the doping of 0.23  we find our topology change to occur at.  On the other hand, our doping value does correspond to the one  where collective effects  were found to disappear in the in-plane infrared conductivity \cite{TIMUSK}.  And it is close to the value where c-axis transport indicates that T$_c$ and the pseudogap temperature T* merge \cite{KRUSIN}.
 
One might also naively expect that a change of topology would affect transport properties, especially the Hall effect.  The dependence of the Hall effect on doping, though, is smooth in this range of doping \cite{RAFFYPRB}.  This is not a surprise, though, since the Fermi velocity of the antibonding sheet is quite low in the ($\pi$,0) region of the Brillouin zone.  Tight binding simulations we have made of the Hall effect find a hole-like behavior for dopings far in excess of where the topology change occurs. However, the presence of the van Hove singularity in the antibonding sheet very close to the chemical potential could have consequences for those properties which are sensitive to the breaking of particle-hole symmetry.

In summary, we report the observation of a change in the Fermi surface topology in the heavily overdoped regime of Bi$_2$Sr$_2$CaCu$_2$O$_{8+\delta}$. In overdoped samples, at dopings slightly below those corresponding to T$_c$=55K, the antibonding band changes from hole-like to electron-like. The bonding band remains hole-like over the whole doping range. This finding has potential consequences for theories of high temperature superconductivity as well as the interpretation of transport measurements in this doping range.

We acknowledge useful remarks by Joe Orenstein and Andrey Chubukov and the hospitality of the Aspen Center for Physics, where parts of this work were discussed and written up. This work was supported by the NSF DMR 9974401 and the US DOE, Office of Science, under Contract No. W-31-109-ENG-38. The Ames Laboratory is operated for the US DOE by Iowa State University under Contract No. W-7405-ENG-82. The Synchrotron Radiation Center is supported by NSF DMR 9212658.

\end{document}